\documentclass[submission,copyright,creativecommons]{eptcs}
\providecommand{\event}{ICLP 2026} 

\usepackage{tikz}
\usepackage{pgfplots}
\pgfplotsset{compat=1.18}
\usepgfplotslibrary{statistics}
\usetikzlibrary{shapes.geometric, positioning, arrows.meta, fit, backgrounds, calc}

\usepackage{amssymb}
\usepackage{booktabs}
\usepackage{multirow}

\newcommand{\aspcode}[1]{\texttt{{#1}}}
\newcommand{\clingo}[0]{\texttt{clingo}}

\newcommand{\avgdur}[0]{\text{aveg\_dur}}
\newcommand{\demord}[0]{\text{dem\_ord}}
\newcommand{\roomexam}[0]{\text{room\_exam}}
\newcommand{\roomcap}[0]{\text{room\_cap}}

\newcommand{\travtime}[0]{\text{trav\_time}}
\newcommand{\arrtime}[0]{\text{arr\_time}}
\newcommand{\queue}[0]{\text{queue\_at}}
\newcommand{\queuecur}[0]{\text{queue\_{0}}}
\newcommand{\hist}[0]{\text{hist}}
\newcommand{\timeh}[0]{\ensuremath{\{0, \dots, H\}}}

\usepackage{subcaption}

\usepackage{iftex}

\ifpdf
  \usepackage{underscore}        
  \usepackage[T1]{fontenc}        
  \usepackage{amsmath}            
\else
  \usepackage{breakurl}           
\fi
\usepackage{graphicx}
\title{Walk-In Multi-Stage Patient Flow Scheduling: An ASP Model with DES-Based Evaluation}
\author{
Ngoc-Mai Pham\thanks{These authors contributed equally to this work.}
\qquad\qquad Trang-Linh Nguyen\footnotemark[1]
\qquad\qquad Thi-Hai-Yen Vuong
\institute{Knowledge Technology Laboratory, Faculty of Information Technology, \\VNU University of Engineering and Technology, Hanoi, Vietnam}
\and
Ha-Thanh Nguyen
\institute{College of Engineering \& Computer Science, VinUniversity, Hanoi, Vietnam\\National Institute of Informatics, Tokyo, Japan}
\and
Van-Giang Trinh\thanks{Corresponding author: van-giang.trinh@hcmut.edu.vn}
\institute{Faculty of Computer Science and Engineering, Ho Chi Minh City University \\of Technology (HCMUT), VNU-HCM, Ho Chi Minh City, Vietnam}
}
\def\titlerunning{Walk-In Multi-Stage Patient Flow Scheduling}
\def\authorrunning{N.-M. Pham, T.-L. Nguyen, T.H.-Y. Vuong, H.-T. Nguyen, \& V.-G. Trinh}
\def\authorrunning{N.-M. Pham, T.-L. Nguyen, T.H.-Y. Vuong, H.-T. Nguyen, \& V.-G. Trinh}
\def\copyrightholders{N.-M. Pham, T.-L. Nguyen, \\ T.H.-Y. Vuong, H.-T. Nguyen, \& V.-G. Trinh}
\begin{document}
\maketitle

\begin{abstract}
An effective examination and test schedule for patients plays a crucial role in hospital resource management.
In this work, we formulate a new reactive patient-flow scheduling problem in multi-department hospitals where walk-in patients arrive over time and each patient requires multiple examinations per visit.
Upon each arrival, the scheduler computes a feasible examination pathway---both the sequence of examinations and the room assignment---for the incoming patient only, while previously scheduled assignments remain fixed.
This process is subject to medical precedence constraints and room capacity limitations.
We model the problem declaratively in Answer Set Programming (ASP) with clingo, and optimize a two-part cost: travel time between consecutive examination locations and queue-induced waiting time, weighted by the duration of the upcoming examination.
To assess robustness under stochastic service times, we propose a Discrete-Event Simulation (DES) evaluation layer and a baseline greedy policy for comparison.
On large-scale synthetic datasets across various capacity regimes and patient loads, the ASP approach reduces median stay time and increases the proportion of zero-wait patients compared to DES-based baselines. 
These improvements are most pronounced under heavy load, while the approach still outperforms baselines across all capacity settings, with smaller gains at higher capacities.
\end{abstract}

\section{Introduction}

An effective examination schedule is crucial for hospital resource management, reducing congestion, balancing workloads, and improving patient satisfaction.
However, overcrowding and long waiting times remain common, especially during peak hours~\cite{Cayirli2003,AJK2017}.
Process improvement methods such as Lean management and value stream mapping have been applied, including in Vietnamese hospitals~\cite{Duong2021}, but they mainly address inefficiencies rather than scheduling decisions.
In practice, scheduling is often manual or rule-based, making it difficult to balance room utilization, adapt exam sequences, and account for inter-department travel~\cite{AD2015}.
Moreover, real-time decision support for patient arrivals is largely absent.

Patient scheduling and patient-flow management in healthcare systems have been extensively studied using queueing theory, simulation, and optimization methods to address uncertainty in service times and limited medical resources~\cite{Jun1999,Papiya2014,Leeftink2017}.
More recent studies extend these approaches through decision-support systems for outpatient appointment scheduling and hospital resource allocation~\cite{CSWL2024,NLGFLGYG2024,YSP2024,Bigharaz2025}.
Optimization techniques such as Mixed-Integer Linear Programming (MILP), metaheuristics, and stochastic programming have been widely applied to improve patient access and resource utilization~\cite{Gupta2008,BMHF2019,Kuo2019,QWWY2019,FTYGC2021,ASPT2022,YSP2024}.
A few learning-based methods have also been proposed~\cite{Ala2022}.

More recently, Answer Set Programming (ASP)~\cite{GKKS2012} has emerged as a promising approach for solving complex healthcare scheduling problems due to its ability to model rich constraints declaratively~\cite{ADM2018,CGNR2023,Kanias2023,CGMMP2024,VMCPP2025}.
While the ASP framework in~\cite{VMCPP2025} addresses multi-session scheduling, it optimizes inter-session intervals and home-to-clinic travel rather than multi-stage patient flows and intra-hospital routing within a single visit. 
Furthermore, its evaluation relies solely on static ASP solving, lacking dynamic simulation to assess real-world operational performance.

In many hospitals in developing countries such as Vietnam, patient visits often involve multiple examinations across different departments without pre-booked appointments~\cite{nguyen2018waiting,Quan2023,Tho2017}.
Unlike appoint\-ment-based systems, patients typically receive a list of required tests after an initial consultation and proceed in a walk-in manner~\cite{Quan2023}.
This setting creates a complex patient-flow problem: patients must follow partially ordered examinations, compete for limited resources, and travel between departments.
Existing healthcare scheduling research has mainly focused on appointment-based or single-resource models using queueing, simulation, or optimization techniques~\cite{BMHF2019,Kuiper2021,YSP2024}.
These approaches typically assume predefined schedules or fixed treatment sequences and rarely incorporate intra-hospital patient movement into the optimization process~\cite{Morikawa2017,Pan2020}.

In this work, we formulate a multi-stage patient flow scheduling problem in which each walk-in patient may require multiple examinations during a single visit.
The problem involves coordinating examination sequences across departments under room capacity constraints and inter-department travel-time costs.
Unlike traditional appointment scheduling, this setting introduces additional challenges: medical precedence constraints, dynamic competition for shared resources, and the impact of patient movement on system efficiency.
The scheduler must therefore jointly determine the examination order and room assignment while minimizing waiting time and unnecessary movement.

This problem differs from existing models by jointly optimizing sequencing, resource allocation, and spatial routing under dynamic arrivals, which are typically treated separately~\cite{Munavalli2020,YSP2024}.
Stochastic programming approaches~\cite{YSP2024} optimize multi-resource allocation and care sequencing under uncertainty, but are designed for appointment-based settings with predefined arrivals, and do not explicitly handle walk-in patients, spatial factors, or real-time adaptation.
Similarly, integrated patient scheduling models~\cite{Munavalli2020} construct fixed care pathways at registration to minimize waiting time and stay, but focus primarily on temporal aspects and overlook travel distance, while relying on static or manually updated system states.
In contrast, our approach unifies these aspects within a reactive framework supporting dynamic arrivals.

We employ ASP to compute patient pathways while minimizing waiting time and travel time.
The model captures both hard constraints (e.g., precedence and capacity) and soft constraints for optimization.
Due to uncertainties in arrivals and service durations, schedules are evaluated using Discrete Event Simulation (DES) under stochastic conditions.
We also compare against representative greedy scheduling strategies used in practice.
By integrating declarative optimization with simulation-based evaluation, our approach provides a flexible decision-support framework for walk-in healthcare environments.

To sum up, our main contributions include:
\begin{itemize}
\itemsep0em
    \item A novel reactive scheduling formulation for multi-stage walk-in patient pathways, jointly deciding exam order and room assignment under precedence, capacity, and travel constraints.
    \item An ASP encoding with lexicographic optimization that prioritizes minimizing waiting time over travel cost.
    \item A DES-based evaluation framework capturing stochastic service dynamics and enabling realistic performance assessment.
    \item Large-scale synthetic experiments showing consistent reductions in patient stay time and higher zero-wait rates compared to DES baselines, especially under heavy workloads.
\end{itemize}

The remainder of the paper is organized as follows.
Section~\ref{sec:problem-formulation} presents the problem formulation, Section~\ref{sec:encoding} describes the ASP encoding, Section~\ref{sec:experiments} reports experimental results and discusses the findings and limitations.
Finally, Section~\ref{sec:conclusion} concludes the paper and outlines directions for future work.

\section{Problem Formulation}\label{sec:problem-formulation}

\subsection{System Description}

\begin{figure}[!ht]
\centering
\begin{tikzpicture}[scale=0.75,
  base/.style={
    draw, rectangle,
    minimum width=3.0cm, minimum height=0.85cm,
    align=center, font=\small, line width=0.8pt
  },
  trigger/.style={
    draw, diamond, aspect=2.2, inner sep=2pt,
    align=center, font=\small, line width=0.8pt
  },
  arrow/.style={-Stealth, line width=0.8pt},
  line/.style={line width=0.8pt},
  seclabel/.style={font=\small\bfseries}
]

\node[base] at (0,  1.4) (baseinfo) {Hospital Configuration};
\node[base] at (0,  0.0) (status)    {Current System State};
\node[seclabel] at (0, 2.3) (inputlabel) {INPUT};

\begin{scope}[on background layer]
  \node[draw, dashed, rounded corners=8pt, inner sep=10pt,
        fit=(inputlabel)(baseinfo)(status)] (inputgroup) {};
\end{scope}

\node[base, minimum width=2.8cm, minimum height=1.5cm] at (5.5, 0) (solver)
      {\textbf{Scheduling Solver}};

\node[trigger] at (5.5, 2.5) (arrival) {New Arrival};

\node[base] at (10.5, 0) (examseq) {Patient Pathway};
\node[seclabel] at (10.5, 0.9) (outputlabel) {OUTPUT};

\begin{scope}[on background layer]
  \node[draw, dashed, rounded corners=8pt, inner sep=10pt,
        fit=(outputlabel)(examseq)] (outputgroup) {};
\end{scope}

\coordinate (merge) at (3.4, 0);

\draw[line] (baseinfo.east) -- ++(0.55,0) |- (merge);
\draw[line] (status.east)    --              (merge);

\draw[arrow] (merge) -- (solver.west);

\draw[arrow] (arrival.south)
    -- node[right, font=\small, xshift=3pt] {\textit{trigger}}
    (solver.north);

\draw[arrow] (solver.east) -- (examseq.west);

\draw[arrow]
    (outputgroup.south)
    -- ++(0, -0.9)
    -- node[below, font=\small] {\textit{feedback}}
        ($(inputgroup.south) + (0, -0.9)$)
    -- (inputgroup.south);

\end{tikzpicture}

\caption{Reactive scheduling framework for multi-stage patient pathways}
\label{fig:schedulingframework}
\end{figure}
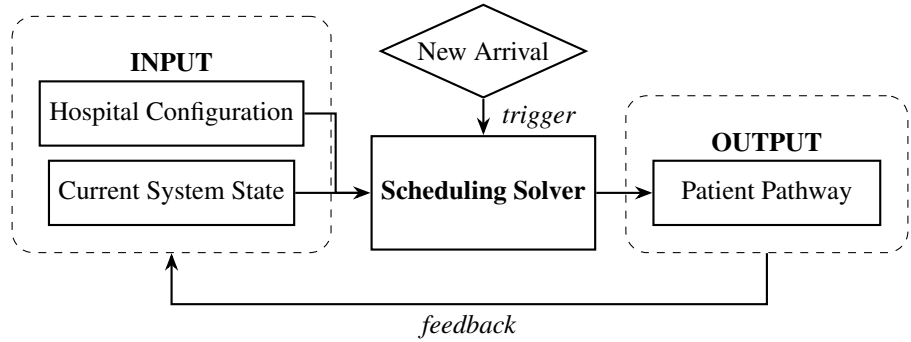

We consider a dynamic patient-flow scheduling problem arising in walk-in hospital environments.
Patients may require multiple examinations during a single visit, and these examinations are performed in different departments of the hospital.
Figure~\ref{fig:schedulingframework} illustrates the overall workflow of the proposed scheduling system.
The system operates in a reactive manner: whenever a new patient arrives, the scheduler computes a feasible examination pathway based on the current hospital state.

The system receives two main input categories, along with the examination requirements of the incoming patient.
The \textit{hospital configuration} describes the static structure of the hospital, including the set of examination types, examination rooms, room capacities, average examination durations, travel times between examination areas, and medical precedence constraints.
The \textit{current system state} summarizes the operational status of the hospital at the decision time, including the estimated queues at examination rooms and the activities of previously scheduled patients.
In addition, the \textit{examination requirements} of the incoming patient define the set of required examinations.

Based on this input, the solver schedules exam sequences and room assignments to minimize wait and travel times for the incoming patient.
Each new pathway updates the system state, allowing the scheduler to dynamically adapt to evolving hospital conditions as new patients arrive.

In the following subsection, we formally define the scheduling problem solved at each arrival event.

\subsection{Mathematical Formulation}\label{subsec:mathematical-formulation}

\paragraph{Sets}

\begin{itemize}
\itemsep0em
\item $E = \{e_1, \dots, e_{|E|}\}$: set of examination types
\item $P = \{p_1, \dots, p_{|P|}\}$: set of patients scheduled, used only to define the parameter $hist$ afterwards
\item $EL = \{l_1,\dots,l_{|EL|}\}$: set of examination locations (rooms)
\item $E_{np} \subseteq E$: set of examinations required by the incoming patient $np$
\item $\timeh$: hospital operating horizon, considering discrete time (in minutes) where $H \geq 0$ is the total active time of the hospital
\end{itemize}

\paragraph{Parameters}

\begin{itemize}
\itemsep0em
\item $\avgdur: E \rightarrow \mathbb{N}$: average duration of each examination (in minutes, rounded to the nearest integer)
\item $\demord: E \rightarrow \mathbb{N}$: precedence level of each examination
\item $\roomexam: EL \rightarrow E$: surjective mapping from rooms to examination types
\item $\roomcap: EL \rightarrow \mathbb{N}$: capacity of each examination room
\item $\travtime: EL \times EL \rightarrow \mathbb{N}$: travel time between examination locations (in minutes, rounded to the nearest integer)
\item $\queuecur: EL \rightarrow \mathbb{N}$: number of patients currently waiting at a room at the decision time
\item $\hist: EL \times \timeh \rightarrow \mathbb{N}$: total number of patients scheduled to arrive at room $r$ within the timeframe $\{0, \dots, t\}$ ($t \in \timeh$) based on prior scheduling decisions
\end{itemize}

\paragraph{Decision variables}

The scheduler constructs an ordered pathway
\[
\Pi = \big((e_{i_{1}},l_{j_{1}}), (e_{i_{2}},l_{j_{2}}), \dots, (e_{i_{|E_{np}|}},l_{j_{|E_{np}|}})\big),
\]
where $\forall k \in \{1, \dots, |E_{np}|\}$, $e_{i_{k}} \in E_{np}$ and $l_{j_{k}} \in EL$.

\paragraph{Derived quantities}

\begin{itemize}
\itemsep0em
\item $\arrtime: E_{np} \rightarrow \timeh$: estimated arrival time for an examination of the incoming patient
\[
\arrtime(e_{i_{k+1}})=\arrtime(e_{i_k})+\avgdur(e_{i_k})+\travtime(l_{j_k},l_{j_{j_k}})
\]
\item $\queue: EL \times \timeh \rightarrow \mathbb{N}$: estimated queue length at room $l \in EL$ at time $t \in \timeh$, derived from $\queuecur$ and $\hist$. More specifically,
\[
\queue(l,t) = \max\left(0, \queuecur(l) + \hist(l,t) - \left\lfloor \frac{t\times\roomcap(l)}{\avgdur(\roomexam(l))}\right\rfloor\right)
\]

\end{itemize}

\paragraph{Feasibility constraints}

The pathway $\Pi$ must satisfy the following conditions:
\begin{enumerate}
\itemsep0em
\item Each required examination appears exactly once:
\[
\forall e\in E_{np}, \exists!k \in \{1, \dots, |E_{np}|\} | e_{i_k} = e.
\]
\item Each assigned room must be compatible with the corresponding examination:
\[
\roomexam(l_{j_{k}}) = e_{i_{k}}, \quad \forall k \in \{1, \dots, |E_{np}|\}.
\]

\item Medical precedence constraints must be respected:
if $\demord(e_{i_{k}}) < \demord(e_{i_{j}})$ and both $e_{i_{k}}, e_{i_{j}} \in E_{np}$, then $e_{i_{k}}$ must appear before $e_{i_{j}}$ in the pathway $\Pi$, i.e., $k < j$.

\end{enumerate}

\paragraph{Objective}

For the computed pathway, we consider two costs.

\textbf{Travel cost.}
\[
Z_{travel} = \sum_{k = 1}^{|E_{np}| - 1}\travtime(l_{j_{k}}, l_{j_{k + 1}}).
\]

\textbf{Waiting cost.}
\[
Z_{wait} = \sum_{k = 1}^{|E_{np}|}\frac{\queue(l_{j_{k}}, \arrtime(e_{i_{k}})) \times \avgdur(e_{i_{k}})}{\roomcap(l_{j_k})}.
\] 
The optimization is lexicographic: the model first minimizes $Z_{wait}$ and then minimizes $Z_{travel}$.

\paragraph{Reactive update}

To formalize the reactive update process, suppose that we have calculated the optimal pathway $\Pi^{(m)}$ for the current patient $p_m$. This computed pathway is then added to the current schedule to update the system state. Specifically, the updated state becomes the input for the next patient arrival, $p_{m+1}$, by updating the historical schedule parameter, $\hist$, as follows:
\[
\hist^{m+1}(l,t)=\hist^{m}(l,t)+|\{k\in\{1,\dots,|E_np|\}|l_{j_{k}}=l\land \arrtime(e_{i_k}) \le t\}|
\]

\section{ASP Encoding}\label{sec:encoding}

In this section, we present an Answer Set Programming (ASP) model for the proposed scheduling problem.
Before describing the encoding, we briefly discuss why ASP is a suitable framework for this task.

\subsection{Why ASP?}

The proposed scheduling problem combines multiple decision dimensions, including examination sequencing, room assignment, precedence constraints, room capacities, and patient travel time.
ASP provides a natural declarative framework for modeling such combinatorial problems.
Complex hospital workflow rules can be expressed directly as logical constraints, allowing the model to remain compact and easily adaptable when policies or resource configurations change.

ASP is particularly suitable for jointly optimizing multiple objectives.
In our formulation, the solver simultaneously determines the order of examinations and the assignment of examination rooms while minimizing waiting time and travel time.
In addition, ASP is well-suited to reactive scheduling in walk-in environments: when a new patient arrives, the solver can be re-triggered with updated facts representing the current hospital state.
These features make ASP a flexible and transparent framework for solving the proposed reactive multi-stage patient-flow scheduling problem.

\subsection{Representation of Hospital State}

The main atoms used to represent the hospital state are as follows:
\begin{itemize}
\itemsep0em

\item \aspcode{exam(ExamID, Dur)}: examination type \aspcode{ExamID} has duration \aspcode{Dur}

\item \aspcode{examLoc(ExlID, ExamID, Cap)}: room \aspcode{ExlID} supports examination \aspcode{ExamID} with capacity \aspcode{Cap}

\item \aspcode{phase(ExamID, Order)}: examination type \aspcode{ExamID} has precedence level \aspcode{Order}

\item \aspcode{travel(ExA, ExB, Time)}: defines the travel time between \aspcode{ExA} and \aspcode{ExB}.
While the mathematical formulation defines travel costs between specific physical rooms $\travtime(l_i,l_j)$, this ASP encoding abstracts distances to the exam-type level.
This abstraction reflects practical hospital layouts, where rooms providing identical services are typically close together, while reducing encoding complexity

\item \aspcode{needs_exam(examID)}: the examinations required by the incoming patient \aspcode{examID}

\item \aspcode{assign(E, R, S)}: a scheduling decision, assigning examination \aspcode{E} to room \aspcode{R} and step \aspcode{S}

\item \aspcode{patient_act(P, R, T_arr)}: represents the arrival times of patients already scheduled in the system. It is used to calculate $\hist(l, t)$.
The atom only needs arrival times to count the incoming patients because the model estimates processed patients using throughput \aspcode{T_arr * Cap / Dur} instead of exact finish times as discussed later in Section~\ref{subsec:Optimization-Criteria}.

\item \aspcode{queue_at(R, T_arr, N)}: indicates that \aspcode{N} patients are waiting at room \aspcode{R} at time \aspcode{T_arr}. It represents $\queue(l, t)$ and $\queuecur(l)$ in case $t = 0$.
\end{itemize}

Note that every time the system is run for a new patient, facts in the form of \aspcode{needs_exam(examID)}, \aspcode{queue_at(R, T_arr, N)}, and \aspcode{patient_act(P, R, T_arr)} are provided to the solver.

\subsection{Hard Constraints}

The ASP encoding presented in this section is logically equivalent to the mathematical formulation in \ref{subsec:mathematical-formulation}.
While the mathematical formulation uses a positional index $k \in \{1, \dots, |E_{np}|\}$ to represent the examination order, this index must be explicitly grounded in ASP.
We therefore compute the number of required exams $N = |E_{np}|$ using the \aspcode{\#count} aggregate and generate a finite set of steps \aspcode{step(1..N)}. 
Note that, $N$ can be passed to \clingo{} as a constant value, without using the \aspcode{\#count} aggregate.

\begin{small}
\begin{verbatim}
    last_step(N) :- N = #count{E : needs_exam(E)}.
    step(1..N) :- last_step(N).
\end{verbatim}
\end{small}

The hard constraints mentioned in Section~\ref{subsec:mathematical-formulation} are encoded in ASP as follows:

\begin{small}
\begin{verbatim} 
    1 { assign(E, R, S) : examLoc(R, E, _), step(S) } 1 :- needs_exam(E).
    :- assign(E1, _, S), assign(E2, _, S), E1 != E2.
    :- assign(E1, _, S1), assign(E2, _, S2),
       phase(E1, O1), phase(E2, O2), O1 < O2, S1 >= S2.
\end{verbatim}
\end{small}

The first rule assigns each required exam to exactly one room and step.
The second enforces that at most one exam is performed at each step. 
The third ensures that medical precedence constraints, defined by \aspcode{phase/2}, are respected.

Note that the first two mathematical constraints in Section~\ref{subsec:mathematical-formulation} do not map one-to-one to these ASP rules. 
Specifically, constraint $1$ in Section~\ref{subsec:mathematical-formulation} is guaranteed by the combination of the first two ASP rules, while constraint $2$ is satisfied by the \aspcode{examLoc} atom in the first ASP rule.

\subsection{Optimization Criteria}
\label{subsec:Optimization-Criteria}

The atom \aspcode{queue_at(R, 0, N)} captures the initial queue at each exam location, which is updated at the scheduling time.
The queue at arrival time \aspcode{T_arr} is estimated by combining this initial backlog with the workload induced by previously scheduled patients.
The ASP encoding approximates $\queue(l, t)$ as follows:

\begin{small}
\begin{verbatim}
    has_initial_queue(R) :- queue_at(R, 0, _).
    base_queue(R, Q) :- queue_at(R, 0, Q).
    base_queue(R, 0) :- examLoc(R, _, _), not has_initial_queue(R).
    raw_queue(R, T_arr, Q + T - (T_arr * Cap / Dur)) :- 
        base_queue(R, Q), assign(E, R, S),
        examLoc(R, E, Cap), exam(E, Dur),
        arrival_time(S, T_arr),
        T = #count{P,T_past: patient_act(P, R, T_past, _),T_past > 0,T_past <= T_arr}.
    queue_at(R, T_arr, Raw) :- raw_queue(R, T_arr, Raw), Raw >= 0.
    queue_at(R, T_arr, 0) :- raw_queue(R, T_arr, Raw), Raw < 0.
\end{verbatim}
\end{small}

To reduce grounding complexity, we use a relative time frame where the incoming patient's check-in time is set to \aspcode{checkin(0)}.
Arrival times are recursively defined: the first step starts from reception, while subsequent steps depend on the previous finish time and travel time.

\begin{small}
\begin{verbatim}
    arrival_time(1, T_check + T_walk) :- assign(E, R, 1), checkin(T_check), 
        travel("reception", E, T_walk).
    arrival_time(S + 1, T_fin + T_walk) :- step(S), finish_time(S, T_fin), 
        assign(E1, _, S), assign(E2, _, S + 1), travel(E1, E2, T_walk).
\end{verbatim}
\end{small}

Waiting time is derived from the estimated queue using the room service rate: $T_{wait} = \frac{Q_{est} \times \text{Duration}}{\text{Capacity}}$.
The atom \aspcode{wait_duration(S, T_wait)} is defined to compute the expected delay for each step $S$.
This value subsequently determines the \aspcode{start_time}, calculated as \aspcode{T_start = T_arr + T_wait}, and the \aspcode{finish_time}, calculated as \aspcode{T_fin = T_start + Dur}, at the same step.
These timestamps act as interdependent constraints, where the finish time of step $S$ directly dictates the arrival time of step $S + 1$.

\begin{small}
\begin{verbatim}
    wait_duration(S, (Queue * Dur)/Cap) :- assign(E, R, S),    
        arrival_time(S, T_arr), queue_at(R, T_arr, Queue),  
        exam(E, Dur), examLoc(R, E, Cap).
    start_time(S, T_start) :- arrival_time(S, T_arr), wait_duration(S, T_wait), 
        T_start = T_arr + T_wait.
    finish_time(S, T_fin) :- assign(E, _, S), start_time(S, T_start), 
        exam(E, Dur), T_fin = T_start + Dur.
\end{verbatim}
\end{small}

The optimization is encoded using weak constraints, prioritizing waiting time over travel time (to ensure rapid clinical intervention):

\begin{small}
\begin{verbatim}
    :~ assign(E1, _, S), assign(E2, _, S + 1), travel(E1, E2, T). [T @ 1, S]
    :~ assign(E, R1, S),  wait_duration(S, W), exam(E, Dur). [W @ 2, E, R1]
\end{verbatim}
\end{small}

\section{Experiments}\label{sec:experiments}

\subsection{Experimental Questions}

The experimental study aims to answer the following research questions:
\begin{itemize}
\itemsep0em
\item \textbf{RQ1.} How do different optimization objectives (waiting time, travel time, or both) affect patient flow performance and overall hospital stay?

\item \textbf{RQ2.} Can the proposed ASP formulation reduce patient waiting time and total stay time compared to traditional scheduling strategies?

\item \textbf{RQ3.} How does the performance of the proposed approach change as the number of patients increases under fixed hospital capacity?

\item \textbf{RQ4.} How does the relative performance of the ASP-based approach change as hospital capacity increases under comparable utilization levels?
\end{itemize}

Each research question is addressed by specific experimental analysis presented: RQ1 is investigated in Section~\ref{subsec:Solver-Performance}, while RQ2–RQ4 are jointly investigated through Sections~\ref{subsec:Results} and~\ref{subsec:Scalability-Analysis}. The system implementation and the benchmark reproducibility artifact are available at~\url{https://doi.org/10.5281/zenodo.19398276}.

\subsection{Datasets}
\label{subsec:Datasets}

All experiments use synthetic datasets generated based on operational information collected from hospitals in Vietnam.
The parameters were designed to reflect realistic hospital workflows while allowing controlled evaluation across different system loads.

The simulated hospital operates for a 10-hour shift and accepts walk-in patients during the first $8$ hours.
For each examination type, the total number of requests across all patients is limited to at most $90\%$ of the maximum daily capacity.
This constraint reflects common hospital practices where additional patients may be deferred when the remaining capacity becomes insufficient.

The base configuration includes $16$ examination rooms supporting $10$ exam types, with an average capacity of two patients per room.
Travel time between examination areas is modeled as $5 \pm 2$ minutes, with symmetric travel times between any two areas. 
This assumption is based on the healthcare infrastructure in Vietnam, where patients registered in a specific department are typically required to undergo examinations and tests within that same department.
In most hospitals, examination and testing rooms belonging to the same department are usually located within the same building or nearby facilities, which justifies the relatively short travel time assumption of $5 \pm 2$ minutes.

To evaluate system performance under different workloads, we generated three experimental scenarios with 100, 250, and 400 patients.
These correspond to utilization rates of 20\%, 50\%, and 72\%, respectively.
The utilization rate is defined as the ratio between the total number of examinations required by all patients and the maximum examination capacity of the hospital.

To evaluate scalability, we generated an additional dataset in which the rooms and examination types remain unchanged, but room capacities are increased.
This configuration simulates a larger hospital with higher throughput.
The corresponding patient volumes are scaled to maintain the same utilization levels (20\%, 50\%, and 72\%), resulting in 200, 450, and 650 patients, respectively.

\subsection{Baselines}
\label{subsec:Baselines}
We compare the proposed ASP-based scheduling approach with two baseline strategies that capture two representative operational regimes in Vietnamese hospitals (purely reactive and anticipatory heuristics).

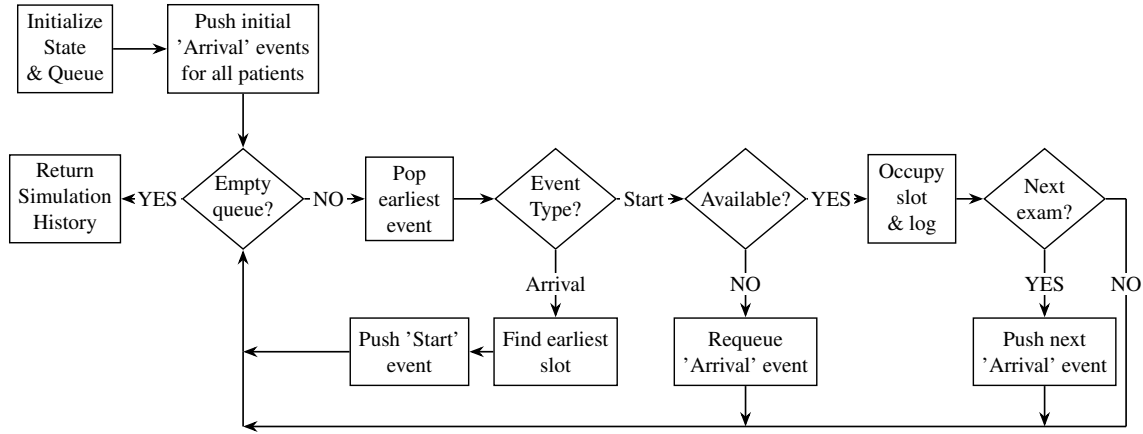
\begin{figure}[!ht]
\centering
\resizebox{0.95\textwidth}{!}{

\begin{tikzpicture}[
    font=\footnotesize,
    process/.style={rectangle, draw, minimum width=1.3cm, minimum height=1.0cm, align=center, line width=0.6pt},
    decision/.style={diamond, aspect=1.2, draw, minimum width=1.8cm, minimum height=1.5cm, align=center, inner sep=0pt, line width=0.6pt},
    arrow/.style={-Stealth, line width=0.7pt},
    node distance=0.8cm and 0.4cm 
]

\node[process] (init) {Initialize\\State\\\& Queue};
\node[process, right=0.8cm of init] (pushinit) {Push initial\\'Arrival' events\\for all patients};

\node[decision, below=of pushinit] (empty) {Empty\\queue?};

\node[process] (return) at (init |- empty) {Return\\Simulation\\History};

\node[process, right=0.9 cm of empty] (pop) {Pop\\earliest\\event};
\node[decision, right=0.6 cm of pop] (type) {Event\\Type?};

\node[decision, right=1cm of type] (verify) {Available?}; 

\node[process, right=0.9 cm of verify] (occupy) {Occupy\\slot\\\& log};
\node[decision, right=of occupy] (next) {Next\\exam?};

\node[process, below=1cm of type] (find) {Find earliest\\slot};
\node[process] (pushstart) at (pop |- find) {Push 'Start'\\event};
\node[process] (requeue) at (verify |- find) {Requeue\\'Arrival' event};
\node[process] (pushnext) at (next |- find) {Push next\\'Arrival' event};

\draw[arrow] (init) -- (pushinit);
\draw[arrow] (pushinit) -- (empty);

\draw[arrow] (empty) -- node[fill=white, inner sep=1.2pt, pos=0.4] {YES} (return);
\draw[arrow] (empty) -- node[fill=white, inner sep=1.2pt, pos=0.4] {NO} (pop);
\draw[arrow] (pop) -- (type);

\draw[arrow] (type) -- node[fill=white, inner sep=1.5pt, pos=0.4] {Start} (verify);

\draw[arrow] (verify) -- node[fill=white, inner sep=1.2pt, pos=0.4] {YES} (occupy);
\draw[arrow] (occupy) -- (next);

\draw[arrow] (type) -- node[fill=white, inner sep=1.5pt] {Arrival} (find);
\draw[arrow] (find) -- (pushstart);

\draw[arrow] (verify) -- node[fill=white, inner sep=1.5pt, pos=0.5] {NO} (requeue);

\draw[arrow] (next) -- node[fill=white, inner sep=1.5pt] (yeslabel) {YES} (pushnext);

\coordinate (bottomwire) at ($(find.south) + (0, -0.6cm)$);
\coordinate (upaxis) at (empty.south);

\coordinate (rightturn) at ($(next.east) + (0.3cm, 0)$);
\draw[line width=0.7pt] (next.east) -- (rightturn);
\draw[line width=0.7pt] (rightturn) -- (rightturn |- bottomwire);

\node[fill=white, inner sep=1.5pt] at (yeslabel -| rightturn) {NO};

\draw[arrow] (rightturn |- bottomwire) -- (upaxis |- bottomwire);
\draw[arrow] (requeue.south) -- (requeue.south |- bottomwire);
\draw[arrow] (pushnext.south) -- (pushnext.south |- bottomwire);
\draw[arrow] (pushstart.west) -- (upaxis |- pushstart.west);

\draw[line width=0.7pt] (upaxis |- bottomwire) -- (upaxis |- pushstart.west);
\draw[arrow] (upaxis |- pushstart.west) -- (empty.south);

\end{tikzpicture}
}
\caption{Patient Flow Simulation.}
\label{fig:DESpatientflow}
\end{figure}

\paragraph{DES (Greedy Baseline):} 
After the clinical consultation, patients follow a predetermined examination sequence.
To determine the examination route, the system reconstructs the current state of examination rooms using Discrete Event Simulation (DES)~\cite{Jun1999}.
Figure~\ref{fig:DESpatientflow} shows the DES procedure used to simulate patient activities and estimate the real-time availability of examination rooms at the time of scheduling.

The simulation considers two event types: Arrival and Start.
Upon an Arrival event, the system identifies the earliest available slot in the corresponding room and schedules a Start event.
When processing a Start event, if the selected slot is unavailable, the event is postponed until a slot becomes free; otherwise, the examination begins and the slot is reserved until completion.
After each examination, the next Arrival event is scheduled, accounting for travel time between departments.
The process continues until all events are handled, producing an execution history that captures room availability over time.

Based on this reconstructed state, the patient is assigned to the room with the earliest available start time.
This strategy follows a greedy first-come-first-served principle and relies solely on the current system state, without considering future congestion.

\paragraph{DES + Future Estimation (DES-F):}
This baseline extends the greedy DES strategy by incorporating a future estimation step.
After reconstructing the current system state through simulation, it predicts future room availability using the same queue-based model as in the ASP approach, ensuring a fair comparison between methods. 
The patient is then assigned to the room with the earliest expected start time.

Together, the two baselines represent two levels of decision support in hospital operations:
\begin{itemize}
\itemsep0em
\item Reactive scheduling based on the current system state
\item Heuristic scheduling with anticipation of future congestion
\end{itemize}

\subsection{Experimental Setup}

All experiments were conducted on the Kaggle platform (Ubuntu 22.04.4 LTS) with an x86\_64 architecture, 4 CPU cores, and 32GB RAM.
Scheduling was performed using \clingo{} 5.8.0.
We tested both \aspcode{--opt-mode=opt}, which returns one optimal model, and \aspcode{--opt-mode=optN}, which enumerates optimal models (selecting the last optimal model).
Since the operational metrics were highly similar, we report results obtained with \aspcode{--opt-mode=optN}.
The reported runtime includes the grounding and solving time under this configuration.

To benchmark the proposed ASP approach against hospital practice, we compare it with two baselines (Section~\ref{subsec:Baselines}): a reactive DES reflecting standard operations, and a DES-F variant incorporating future queue estimation for a fair comparison.

The experiments are designed to evaluate the effectiveness of ASP-based scheduling against heuristic practices commonly used in hospitals.
While optimization-based approaches such as MILP and CP have been studied in the literature, there are limited works addressing the same multi-stage, walk-in setting considered here.
Developing such models solely for comparison would require substantial additional effort and is beyond the scope of this study.

We evaluate the ASP model under four configurations: 
(1) hard constraints only,
(2) minimizing waiting time,
(3) minimizing travel time, and
(4) combined hard and soft constraints.
Performance is assessed using patient-flow metrics, including average waiting time, travel time, and total hospital stay (in minutes).
A patient is considered \textit{zero-wait} if no waiting occurs at any step, and \textit{unfinished} if at least one required examination cannot be completed within the 10-hour window.
We also report the percentage of zero-wait patients and the number of unfinished patients.
To ensure a fair and unbiased performance evaluation, the simulation for these unfinished patients continues beyond the 10-hour operational window until all their examinations are completed.
Consequently, their extended delays are fully incorporated into the reported average wait and stay times.

\subsection{Solver Performance}
\label{subsec:Solver-Performance}
\begin{table}[!ht]
\centering
\small
\caption{Performance comparison of ASP configurations.}
\label{tab:asp-config}
\begin{tabular}{lrrrr}
\toprule
\textbf{Model configuration} & \textbf{Wait (min)} & \textbf{Travel (min)} & \textbf{Stay (min)} & \textbf{Runtime (sec)} \\
\midrule
Hard constraints only & 144.79 & 15.78 & 196.21 & 0.08 \\
Only minimize waiting time & 114.00 & 15.60 & 165.24 & 0.10 \\
Only minimize travel time  & 115.84 & 14.97 & 166.45 & 0.09 \\
Full constraints   & 111.84 & 15.36 & 162.84 & 0.09 \\ 
\bottomrule
\end{tabular}
\end{table}

At each patient arrival, the solver computes an optimal examination pathway based on the current hospital state.
In the initial configuration ($100$, $250$, $450$ patients), the grounding size per request ranges from a few hundred to over $10^5$ rules, with up to $2.5 \times 10^4$ atoms.
Despite this, the total execution time (grounding + solving) remains below $0.1$ seconds.

In the scaled configuration ($250$, $450$, $650$ patients), grounding sizes are comparable, with slightly fewer atoms in some cases, and execution time remains low, averaging below $0.04$ seconds.
The ASP model is thus suitable for reactive scheduling in dynamic hospital environments.
The reduction in generated atoms is driven by the normalization in the waiting time computation. 
As capacity increases, the term Queue × Duration / Capacity compresses the range of waiting times, causing different queue states to map to similar values. 
Since subsequent timestamps depend on these values, this effect propagates and leads to a substantial merging of downstream state representations.

To analyze the effect of optimization objectives, we evaluate four ASP configurations on a fixed scenario with $250$ patients, using the same hospital layout and capacities.
Table~\ref{tab:asp-config} shows that using only hard constraints results in the highest waiting time and total stay.
Minimizing waiting time moderately improves overall performance, indicating that it is the dominant factor in patient flow.
In contrast, optimizing travel time mainly reduces movement but has a smaller effect on total stay.
Combining both objectives achieves the best balance, reducing waiting time and maintaining shorter hospital stays.
\subsection{Results}
\label{subsec:Results}
\begin{figure}[htbp]
\centering
\scalebox{0.78}{
    \begin{subfigure}{0.30\linewidth}
        \centering
        \includegraphics[width=\linewidth]{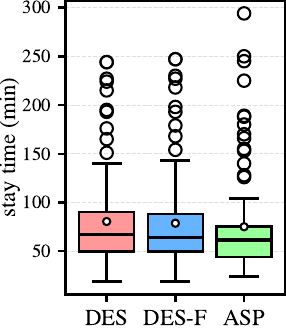}
        \caption{$100$ patients}
        \label{fig:n100}
    \end{subfigure}
    \hspace{0.05\linewidth} 
    \begin{subfigure}{0.30\linewidth}
        \centering
        \includegraphics[width=\linewidth]{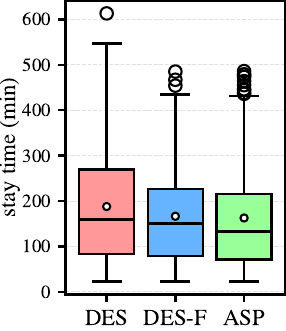}
        \caption{$250$ patients}
        \label{fig:n250}
    \end{subfigure}
    \hspace{0.05\linewidth}
    \begin{subfigure}{0.30\linewidth}
        \centering
        \includegraphics[width=\linewidth]{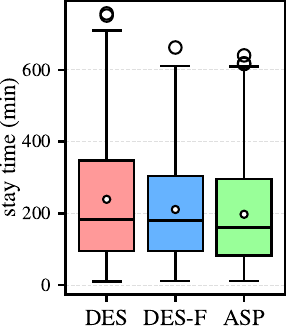}
        \caption{$400$ patients}
        \label{fig:n400}
    \end{subfigure}
}
    \caption{Comparison of Stay Time distributions across different patient scenarios.}
    \label{fig:comparisonresult}
\end{figure}

Figure~\ref{fig:comparisonresult} compares the proposed ASP-based approach with DES-based baselines under low-capacity settings across different patient volumes.
In all scenarios, ASP consistently achieves lower median stay times, and this advantage remains stable as the system scales from $100$ to $400$ patients.

\begin{table}[!ht]
\caption{Performance metrics comparison.}
    \label{tab:metricscomparison}
    \centering
    \small
    \scalebox{0.9}{
    \begin{tabular}{ll ccc ccc}
        \toprule
        \multirow{2}{*}{\textbf{Metric}} & \multirow{2}{*}{\textbf{\shortstack{Scheduling \\ Method}}} & \multicolumn{3}{c}{\textbf{Low capacity scenario}} & \multicolumn{3}{c}{\textbf{High capacity scenario}} \\
        \cmidrule(lr){3-5} \cmidrule(l){6-8} 
        & & \textbf{$N = 100$} & \textbf{$N = 250$} & \textbf{$N = 400$} & \textbf{$N = 200$} & \textbf{$N = 450$} & \textbf{$N = 650$} \\
        \midrule
        
        \multirow{3}{*}{\textbf{\shortstack{Avg. Travel \\ Time (min)}}} 
        & DES           & 15.4 & 15.8 & 13.1 & 14.8 & 16.2 & 14.5 \\
        & DES-F         & 15.4 & 15.8 & 13.1 & 14.8 & 16.2 & 14.5 \\
        & \textbf{ASP}  & \textbf{14.8} & \textbf{15.4} & \textbf{13.0} & \textbf{14.5} & \textbf{15.8} & \textbf{14.4}\\
        \midrule
        
        \multirow{3}{*}{\textbf{\shortstack{Avg. Wait \\ Time (min)}}} 
        & DES         & 28.0 & 136.6 & 197.1 & 5.1 & 92.8 & 150.5 \\
        & DES-F       & 26.2 & 115.2 & 168.7 & \textbf{4.2} & \textbf{72.7} & 127.4 \\
        & \textbf{ASP}& \textbf{23.2} & \textbf{111.8} & \textbf{155.5} & 5.9 & 74.9 & \textbf{121.9}\\
        \midrule
        
        \multirow{3}{*}{\textbf{\shortstack{Zero-Wait \\ Patients (\%)}}}
        & DES         & 19.0 & 4.4 & 5.0 & 41.0 & 8.7 & 3.4 \\
        & DES-F       & 20.0 & 6.8 & 6.0 & 47.5 & 10.7 & 5.8\\
        & \textbf{ASP} & \textbf{36.0} & \textbf{8.8} & \textbf{6.5} & \textbf{51.5} & \textbf{11.7} & \textbf{7.4} \\
        \midrule
        
        \multirow{3}{*}{\textbf{\shortstack{Unfinished \\ Patients (unit)}}}
        & DES         & 8 & 75 & 142 & 1 & 114 & 207 \\
        & DES-F       & 8 & \textbf{66} & 128 & 1 & 101 & 192 \\
        & \textbf{ASP} & \textbf{5} & 67 & \textbf{121} & 2 & \textbf{96} & \textbf{183} \\
        \bottomrule
    \end{tabular}}
    \caption*{\footnotesize \textit{Note: Bold values indicate the best performance for each metric.}}
\end{table}

As shown in Table~\ref{tab:metricscomparison}, ASP also improves waiting performance: the proportion of zero-wait patients ranges from $6.5$\% to $36$\%, compared to $4.4$\%-$20$\% for the DES baselines.
This improvement is particularly notable under limited capacity, where system congestion increases and high-quality scheduling decisions become more difficult to identify.

In such constrained settings, the DES methods are less effective due to their reliance on local or future-aware heuristics.
In contrast, ASP can flexibly explore alternative examination sequences and assign patients to rooms that become available earlier.
As a result, approximately $54.8$\%-$63.2$\% of patients complete their examinations earlier under ASP-generated schedules.

Overall, these results show that the ASP approach improves patient stay time and scheduling stability through combinatorial optimization of the entire patient pathway, while DES-based methods may suffer from congestion due to step-by-step greedy decisions.

\subsection{Scalability Analysis}
\label{subsec:Scalability-Analysis}
To evaluate scalability, we increased system capacity in a non-uniform manner: capacities of routine exams (5--10 mins) were only slightly expanded (e.g., +1 slot), while long-duration exams (20--25 mins), such as CT and Endoscopy, were scaled more aggressively (from 1--2 up to 5).
The number of patients was correspondingly increased to $200$, $450$, and $650$ to maintain the utilization levels (Section~\ref{subsec:Datasets}).

This scaling strategy leverages the data generation mechanism: since requests are distributed until capacity limits are reached, increasing capacities for long-duration exams results in datasets dominated by time-consuming procedures.
Consequently, this setting acts as a stress test with heavier workloads, where scheduling long tasks is more complex and prone to congestion. 
The ASP solver's ability to handle such scenarios demonstrates its robustness in demanding operational conditions.

\begin{figure}[htbp]
    \centering
    \scalebox{0.8}{
    \begin{subfigure}{0.30\linewidth}
        \centering
        \includegraphics[width=\linewidth]{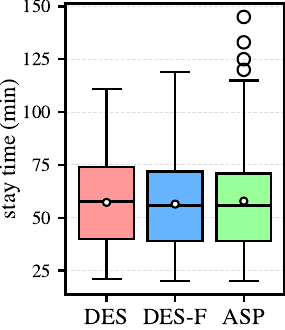}
        \caption{$200$ patients}
        \label{fig:n200}
    \end{subfigure}
    \hspace{0.05\linewidth}
    \begin{subfigure}{0.30\linewidth}
        \centering
        \includegraphics[width=\linewidth]{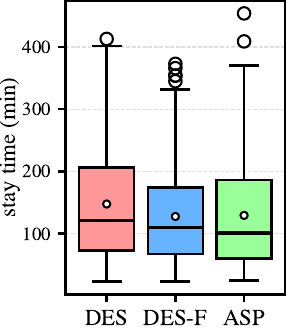}
        \caption{$450$ patients}
        \label{fig:n450}
    \end{subfigure}
    \hspace{0.05\linewidth}
    \begin{subfigure}{0.30\linewidth}
        \centering
        \includegraphics[width=\linewidth]{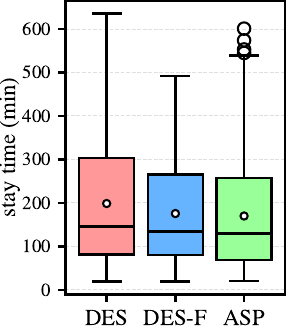}
        \caption{$650$ patients}
        \label{fig:n650}
    \end{subfigure}
    }
    \caption{Comparison of Stay Time distributions across different patient scenarios.}
    \label{fig:comparisonall}
\end{figure}

Overall, while the ASP approach only marginally improves average travel and wait times, the wait time sometimes increases due to strict routing, it consistently optimizes total stay time.
These advantages become more pronounced under higher system loads.

In the low-load scenario ($N = 200$), abundant resources keep waiting times low (around 5 minutes), leading to similar performance across all methods.
As the workload increases ($N = 450$ and $N = 650$), where most exams approach capacity, the strengths of ASP become clear.
In these scenarios, the ASP approach consistently achieves lower median stay times and enables the earliest $25\%$ of patients to finish sooner (Figure~\ref{fig:comparisonall}).
It also improves efficiency, reaching zero-wait rates of 11.7\% and 7.4\% for $N = 450$ and $N = 650$, corresponding to relative gains of 9.3\% and 27.5\% over DES-F, while reducing the number of unfinished patients to 96 and 183, respectively (Table~\ref{tab:metricscomparison}).

In a boxplot, the Interquartile Range (IQR) visually represents the dispersion of the middle 50\% of the data, serving as a robust metric for performance consistency.
In terms of stability, the ASP approach produces a tighter IQR than the standard DES, though slightly wider than DES-F (Figure~\ref{fig:comparisonall}).
This is because ASP relies on local future queue estimates, which may lead to localized bottlenecks. 
In contrast, DES-F follows a fixed examination order, resulting in more stable and consistent patient flow.
Despite this trade-off, ASP still provides a more predictable experience than the standard baseline.
\subsection{Discussion}

\paragraph{Insights}

First, the ASP-based approach consistently improves patient stay time and zero-wait rates compared to the baseline, while remaining computationally feasible even for large-scale scenarios with hundreds of patients.
In low-capacity settings, the solver effectively exploits limited flexibility in room assignment and exam ordering to balance workloads and reduce congestion.

As system capacity increases, however, resource constraints become less binding, yielding many near-equivalent schedules; consequently, the performance gap narrows, as the baseline with future estimation becomes competitive, while the basic baseline remains less effective.

Second, the gains in travel time are relatively small compared to the reductions in waiting time.
This is expected because examination locations are fixed and the number of feasible routing alternatives is limited.
Nevertheless, incorporating travel-time penalties still helps produce smoother patient pathways by discouraging unnecessary movements between departments.

Finally, the integration of ASP-based scheduling with DES evaluation provides a useful methodology for studying healthcare scheduling problems.
While the ASP model optimizes schedules under deterministic assumptions, the DES framework reveals their operational behavior under stochastic service times.

\paragraph{Limitations} Although showing promising results, our evaluation has several limitations.
First, the datasets are synthetic and calibrated from hospital-inspired parameters rather than extracted from operational logs, as the required micro-level data---such as room availability, slot counts, walking times, and examination durations---are rarely publicly available due to privacy restrictions.
Second, the current model assumes First Come, First Serve (FCFS) service within rooms, meaning that patients are processed in the exact order of their arrival without prioritization or interruption. 
Consequently, the model does not yet account for patient acuity, emergency interruptions, or no-shows.
Third, the solver is triggered only when a new patient arrives; re-optimization after each completed examination is left for future work.

\section{Conclusion and Future Work}\label{sec:conclusion}

We introduced a reactive scheduling framework for coordinating multi-stage patient pathways in walk-in hospital environments.
The problem was formulated as a joint optimization of examination ordering and room assignment under precedence, capacity, and travel-time constraints.
We modeled the problem using ASP, enabling a declarative representation of complex scheduling constraints, and evaluated the resulting schedules using DES to capture stochastic operational conditions.

Experiments on large-scale synthetic datasets show that the ASP-based approach consistently improves patient flow compared to heuristic scheduling baselines, reducing median patient stay time and increasing the proportion of zero-wait patients across different capacity regimes.
These results highlight the applicability of ASP for modeling and solving complex scheduling problems involving multi-stage workflows and dynamic system states.

Future work will focus on incorporating real hospital data, extending the model with stochastic arrival processes, and integrating the framework into interactive hospital scheduling systems. 

\bibliographystyle{eptcs}
\bibliography{generic}
\end{document}